\documentclass[journal=cmatex,manuscript=article]{achemso}
\usepackage[usenames,dvipsnames,table]{xcolor}
\usepackage[version=3]{mhchem} %
\usepackage{miller}
\usepackage{subcaption}
\usepackage{multirow}
\usepackage{color}
\usepackage{textgreek}

\graphicspath{{Figures/}}
\usepackage{xspace}
\usepackage{pdfpages}
\usepackage[%
  breaklinks,             %
  bookmarks = false, %
  pdfpagemode   = UseNone,%
  pdfstartview  = FitH,     %
  pdfstartpage  = 1,      %
  colorlinks    = true,   %
]{hyperref}

\newcommand\myshade{85}
\colorlet{mylinkcolor}{YellowOrange}
\colorlet{myurlcolor}{Aquamarine}
\colorlet{mycitecolor}{violet}

\hypersetup{
  linkcolor  = mylinkcolor!\myshade!black,
  citecolor  = mycitecolor!\myshade!black,
  urlcolor   = myurlcolor!\myshade!black,
  colorlinks = true,
}



\author{Bilal Ahmad}
\affiliation{Department of Mechanical Engineering, Texas Tech University, Lubbock, Texas 79409, USA}
\author{Md Salman Rabbi Limon}
\affiliation{Department of Mechanical Engineering, Texas Tech University, Lubbock, Texas 79409, USA}
\author{Zeeshan Ahmad}
\affiliation{Department of Mechanical Engineering, Texas Tech University, Lubbock, Texas 79409, USA}
\email{zeeahmad@ttu.edu}

\title{Modulation of Point Defect Properties Near Surfaces in Metal Halide Perovskites}


\begin{document}

\begin{tocentry}
    \centering
    \includegraphics[width=\textwidth]{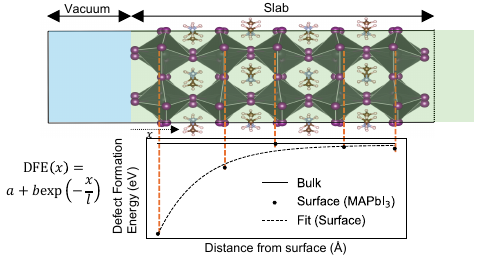}

\begin{center}

\end{center}
\end{tocentry}

\begin{abstract}
It is now widely recognized that surface and interfacial defects exhibit distinct behavior compared to bulk defects in metal halide perovskites. However, the transition from bulk to surface behavior and the spatial extent of the surface's influence are not well understood. 
To address this, we conducted first-principles calculations on iodine vacancies and interstitial defects in methylammonium lead iodide and cesium lead iodide at various depths from the surface, enabling us to map out depth-dependent behavior.
We find that the defect formation energy follows a saturating exponential curve as the defect moves away from the surface to the bulk. 
Using first-principles calculated defect formation energies, we quantify the extent of the surface's influence by calculating the decay length associated with each defect. 
The difference between the surface and bulk defect formation energy is found to be as high as 1.12 eV for the negatively charged iodine vacancy in methylammonium lead iodide, leading to the enrichment of the surface with defects.
Through analysis of defective structures, we find that the differences in the bulk and surface defect properties are a consequence of different bond lengths and in some cases, even changes in bonding and coordination environments. Finally, we determine how the defect transition levels change as a function of the layer index, which could contribute to increased non-radiative recombination.  Our findings pave the way for a systematic treatment of non-radiative losses in perovskite solar cells that incorporate spatially dependent defect densities and transition levels.

\end{abstract}

\section{Introduction}

Metal halide perovskites (MHPs) have emerged as next-generation materials for photovoltaic and optoelectronic applications, demonstrating exceptional power conversion efficiency (PCE) of over 25\%, low-cost solution processing, and optimal band gap~\cite{NRELchart,Huang2017}, making them a promising replacement for traditional silicon-based solar cells. Their structural arrangement enables unique optoelectronic characteristics, including enhanced light absorption, long carrier lifetimes, high carrier mobility, and tolerance to defects~\cite{Kim,Green2014emergence,yooEfficientPerovskiteSolar2021,liChemicallyDiverseMultifunctional2017}.

For a long time, MHPs were considered defect-tolerant materials due to high PCE despite high defect densities introduced during low-temperature processing.  However, recent studies have shown that the defects including iodine vacancies and interstitial defects can act as charge traps or recombination centers, significantly impacting their performance, contradicting the perceived defect tolerance\cite{zhangIodineInterstitialsCause2020,zhangCorrectlyAssessingDefect2020,agiorgousisStrongCovalencyInducedRecombination2014a,alkauskasFirstprinciplesTheoryNonradiative2014}. 
This trapping process promotes a process of recombination of these electrons and holes without the emission of light, resulting in substantial energy losses in devices such as solar cells and light-emitting diodes. In addition, defects that act as recombination centers have low migration energies, suggesting an intricate link between non-radiative recombination processes and hysteresis in MHP solar cells\cite{doi:10.1021/acsenergylett.7b00239}. This insight has led to the conclusion that controlling defect densities may also benefit their operational stability.

Surfaces, interfaces, and grain boundaries contribute disproportionately to the non-radiative recombination processes as pointed out in recent reports\cite{niResolvingSpatialEnergetic2020,meggiolaroFormationSurfaceDefects2019,parkAccumulationDeepTraps2019}. The recombination lifetime at the surface was found to be reduced by two orders of magnitude compared to the bulk~\cite{wuDiscerningSurfaceBulk2016}.  Another experimental study found that the concentration of trap states at the interfaces of MHP solar cells was four orders of magnitude higher than in the bulk~\cite{niResolvingSpatialEnergetic2020}. High surface trap density in MHPs has also been attributed as the cause of narrow band photodetection~\cite{fangHighlyNarrowbandPerovskite2015}. These phenomena have been supported by first-principles calculations showing high defect density at surfaces and interfaces due to reduced defect formation energies (DFEs)~\cite{wooInhomogeneousDefectDistribution2023,ahmadUnderstandingEffectLead2022,meggiolaroFormationSurfaceDefects2019,ambrosioChargeLocalizationTrapping2020,ahmadInterfacesSolidElectrolyte2021}. The high defect density leads to increased non-radiative recombination in these regions. However, currently, we lack an understanding of 1) nature of the transition in defect properties from bulk to the surface, 2) the causes of the low DFEs at surfaces and interfaces, and 3)  changes in other defect properties such as transition levels near surfaces. By understanding near surface defects, chemical treatments that build a protective layer (with optimum depth), can significantly reduce these flaws, hence improving the efficiency and durability of MHP-based devices~\cite{doi:10.1021/acsenergylett.1c02847}.

Here, we uncover the transition in the two properties, DFE and charge transition levels, from the surface to the bulk using first-principles calculations of defects in two MHPs, methylammonium lead iodide (MAPI) and cesium lead iodide (\ce{CsPbI3}). We focus on iodine vacancies and interstitial defects in MHPs which significantly impact non-radiative recombination and ion migration in MHPs. Our calculations of surface slabs allow probing of the nature of defects at different depths from the surface, leading to a functional relationship between defect property and surface depth. Using the DFE profile obtained, we introduce the concept of decay length to determine the spatial extent of the influence of the surface on defects.    We find significant differences in the DFE between bulk and surface defects that may cause defect accumulation or depletion at surfaces, unlike previous studies that found only evidence of defect accumulation~\cite{wooInhomogeneousDefectDistribution2023,meggiolaroFormationSurfaceDefects2019,parkAccumulationDeepTraps2019}.  The low migration energies of these defects will promote rapid rearrangement in response to new surfaces or interfaces. Furthermore, we rationalize the obtained results with changes in the bond lengths and bonding environment around the defect. For the case of iodine interstitials in \ce{CsPbI3}, the surface completely changes the most stable defect site, leading to differences between the DFE in deeper surface layers and the bulk.
We find that the Pb-Pb distance is a good descriptor for explaining the DFE versus surface depth trends observed for iodine vacancies. For iodine interstitials, we find that Pb-I bond lengths explain the convergence of the DFE towards the bulk values.
Besides DFE, we find that the charge transition levels of defects can differ by $\sim 0.4$ eV between the surface and the bulk, which may increase or decrease the non-radiative recombination rate. We emphasize that both DFE and transition levels are important to determine the net effect on non-radiative losses in semiconductors.

\section{Methods}
We used density functional theory (DFT)  as implemented in the Quantum Espresso code~\cite{giannozziQUANTUMESPRESSOModular2009a,giannozziAdvancedCapabilitiesMaterials2017a} to perform calculations on  MAPI and CsPbI$_{3}$ MHPs. The Perdew-Burke-Ernzerhof (PBE) exchange-correlation functional~\cite{perdewGeneralizedGradientApproximation1996} was used together with the norm-conserving pseudopotentials from the Pseudo Dojo library~\cite{vansettenPseudoDojoTrainingGrading2018}. While hybrid functionals may yield more accurate defect properties, their computational cost is prohibitively high for the large surface slabs considered in this study. Our primary objective is to investigate the differences between surface and bulk defects and elucidate the transition between them by probing different depths from the surface. 
We incorporated DFT-D3 dispersion corrections using the scheme proposed by \citealt{Grimme2010} into our computational methodology to account for van der Waals dispersion forces in MHPs. 
We used a 2×2×2 k-point grid for the bulk and a 3×3×1 k-point grid for slabs, to accurately represent the electronic structures. 
Convergence criteria for self-consistency in the calculation of  energy and forces were set to $10^{-4}$ Ry and $10^{-3}$ Ry/Bohr, respectively. We set the energy cutoffs for wave functions to 75 Ry for CsPbI$_{3}$ and 80 Ry for MAPI. The energy cutoffs for electron density were 300 Ry for CsPbI$_{3}$ and 320 Ry for MAPI. We considered the (001) surface for our calculations for both MAPI and CsPbI$_3$. Vacuum of length 10 {\AA} was used on each side of the slabs. Dipole correction was applied to all surface slab calculations.

We performed simulations of iodine vacancy and interstitial defects in different charge states (+1, -1, 0) in CsPbI$_{3}$ and MAPI since these defects have been established as potential non-radiative recombination centers in MHPs~\cite{zhangIodineInterstitialsCause2020,meggiolaroIodineChemistryDetermines2018,agiorgousisStrongCovalencyInducedRecombination2014a}. The  Kroger-Vink notation is followed to represent iodine vacancy in +1, -1, 0 charge states as $\mathrm{V}_{\mathrm{I}}^\bullet$, $\mathrm{V}_{\mathrm{I}}^{{\displaystyle \prime }}$, $\mathrm{V}_{\mathrm{I}}^{\times}$ and iodine interstitials in +1, -1, 0 charge state as  $\mathrm{I}_\mathrm{i}^{\bullet}$, $\mathrm{I}_{\mathrm{i}}^{{\displaystyle \prime }}$, $\mathrm{I}_{\mathrm{i}}^{\times}$ respectively. The defect formation energies (DFE) have been calculated using the expression~\cite{vandewalleFirstprinciplesCalculationsDefects2004,freysoldtFirstprinciplesCalculationsPoint2014}:
\begin{equation}
E_{\text{f}}(\varepsilon_F, \mu_) = E_d - E_p + n \mu + q(\varepsilon_{\text{VBM}} + \varepsilon_F) + \Delta_{\text{corr}},
\end{equation}
where $E_d$ and $E_p$ represent the total energy of the defective and pristine structures (bulk supercell or slab), respectively. $n$ is the number of iodine atoms added or removed from the pristine structure  to create the defect. $\mu$ is the chemical potential of iodine, $q$ is the charge on the defect, $\varepsilon_{\text{VBM}}$ is the valence band maximum (VBM) energy of the pristine structure, $\varepsilon_{\text{F}}$ is the Fermi energy level relative to the VBM of the pristine structure, and $\Delta_{\text{corr}}$ accounts for the correction terms for charged defect calculations. We used the thermodynamic stability diagrams to calculate the chemical potentials of iodine. The secondary phases considered for MAPI were MAI and PbI$_{2}$, whereas for CsPbI$_{3}$, CsI and PbI$_{2}$ were considered as secondary phases. We considered Pb rich conditions with PbI$_{2}$ excess to calculate the chemical potentials. 
The values of $\Delta_{\text{corr}}$ for both MAPI and CsPbI$_3$ are provided in Table S1. We used the method proposed by \citeauthor{freysoldtFirstprinciplesCalculationsPoint2014}\cite{freysoldtFullyInitioFiniteSize2009,freysoldt2018first} to calculate the charge correction terms for bulk and slabs.
The sxdefectalign\cite{freysoldtFirstprinciplesCalculationsPoint2014} code was used for the charge correction terms in bulk supercell (2×2×1 for CsPbI$_{3}$ and $\sqrt{2}$×$\sqrt{2}$×1 for MAPI ) while the sxdefectalign2d\cite{freysoldt2018first} code was used for the correction terms on the surfaces with the dielectric constants of 27.5 and 18 for MAPI\cite{brivioStructuralElectronicProperties2013} and CsPbI$_{3}$\cite{doi:10.1021/acsenergylett.0c02482}, respectively. The lattice parameters for MAPI and \ce{CsPbI3} unit cells are provided in Table S2.

The transition level $\varepsilon_F(q_1/q_2)$ is defined as that value of Fermi energy position where the DFE for the charge states \textit{$q_1$}\ and \textit{$q_2$}\ are equal. It was calculated using the formula~\cite{freysoldtFirstprinciplesCalculationsPoint2014}:
\begin{equation}
\varepsilon_F(q_1/q_2) = \frac{[E_{\text{f}}(\varepsilon_F=0, \mu)]_{q_1} - [E_{\text{f}}(\varepsilon_F=0, \mu)]_{q_2}}{q_2 - q_1},
\end{equation}
where ${[E_{\text{f}}(\varepsilon_F=0, \mu)]}_{q}$ is the DFE of the defect in charge state $q$  corresponding to Fermi energy level equal to VBM ($\varepsilon_F=0$). This transition level determines whether a trap acts as a deep or shallow trap depending on its position within the band gap. If the transition level is closer to the middle of the band gap, the trap state acts as a deep trap, while if the transition level is closer to the band edges, it acts as a shallow trap.

\section{Results and Discussion}
\subsection{Structural analysis}
\begin{figure}[htbp]
    \centering
    \includegraphics[width=\textwidth]{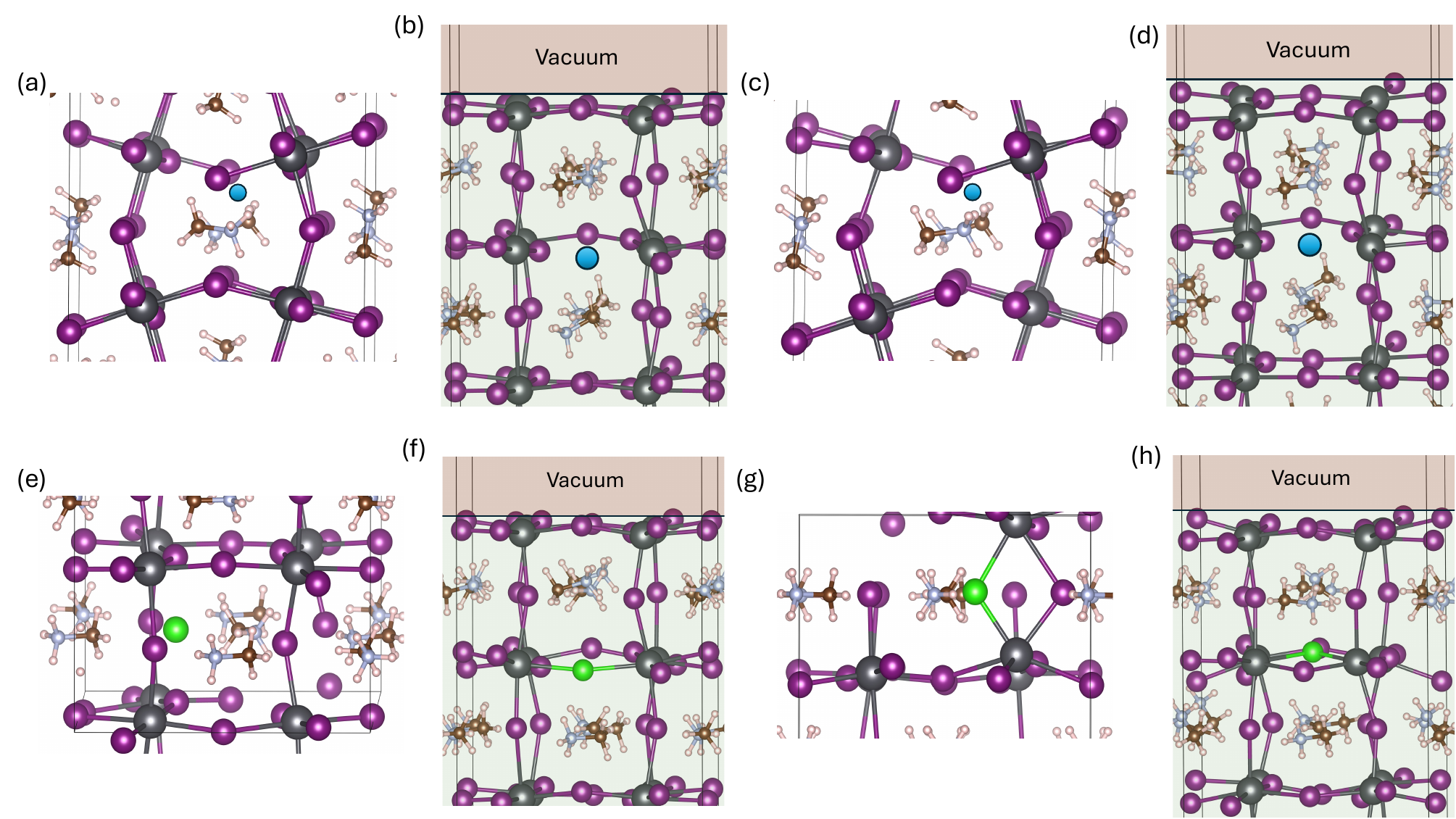}
    \caption{Comparison of atomic structure around iodine vacancies and interstitials in the bulk and surface of MAPI. The surface defects shown are located in the second layer away from the vacuum. The iodine vacancy position is shown using a blue circle and iodine interstitials are shown in green.   The structures depict 
    (a) $\mathrm{V}_{\mathrm{I}}^\bullet$ in the bulk, (b) $\mathrm{V}_{\mathrm{I}}^\bullet$ at the surface, (c) $\mathrm{V}_{\mathrm{I}}^{{\displaystyle \prime }}$ in the bulk, (d) $\mathrm{V}_{\mathrm{I}}^{{\displaystyle \prime }}$ at the surface, (e) $\mathrm{I}_\mathrm{i}^{\bullet}$ in the bulk, (f) $\mathrm{I}_\mathrm{i}^{\bullet}$ at the surface, (g) $\mathrm{I}_{\mathrm{i}}^{{\displaystyle \prime }}$ in the bulk, and (h) $\mathrm{I}_{\mathrm{i}}^{{\displaystyle \prime }}$ at the surface. While the neighborhood of the vacancy is similar for the bulk and surfaces, the iodine interstitial structures show differences in bonding environments for +1 charge. In this case, the iodine interstitial forms asymmetric trimer in the bulk but on the surface, it gets bonded to nearby lead atoms. $\mathrm{I}_{\mathrm{i}}^{{\displaystyle \prime }}$ forms a split interstitial both in the bulk and slab.
}
    \label{fig:MAPI_structures.pdf}
\end{figure}

We first investigate the relaxed structures of iodine vacancies and interstitials in the bulk and surface slabs of MAPI and \ce{CsPbI3}. In both MHPs, the defects were placed in the \ce{PbI2} layer for the bulk and the slabs to facilitate direct comparison.
 \autoref{fig:MAPI_structures.pdf} shows the relaxed atomic geometries in the neighborhood of iodine vacancy and interstitial defects in MAPI bulk and surfaces, illustrating the defect's varying bonding environments. The vacancy site is represented with a blue circle in panels a-d while the interstitial iodine atom is represented in green in panels e-h. 
Panels a and b demonstrate the presence of $\mathrm{V}_{\mathrm{I}}^\bullet$ while c and d illustrate the presence of $\mathrm{V}_{\mathrm{I}}^{{\displaystyle \prime }}$ in the bulk and surface, respectively.

We find no significant differences in the bonding environment between the bulk and surface defects in MAPI for iodine vacancies.
For iodine interstitial defects, we relaxed two different starting configurations. In the first configuration, we placed the interstitial iodine atom between two other iodine atoms, and in the second one, we placed it between two methylammonium groups. We found negligible difference in the energetics between these configurations, hence we chose the former configuration for our analysis.
We noticed that the interstitial iodine in +1 charge state forms a asymmetric trimer in the bulk (panel e) but 
 establishes bonds with neighboring lead atoms at the surface (panel f), indicating a change in energetics caused by the relaxation of the surface. $\mathrm{I}_{\mathrm{i}}^{{\displaystyle \prime }}$ in the bulk and surface (panels g and h) both form a split interstitial where two iodine atoms occupy the same site and act as a bridge between two lead atoms.
 Our bulk configurations for both $\mathrm{I}_\mathrm{i}^{\bullet}$ and $\mathrm{I}_{\mathrm{i}}^{{\displaystyle \prime }}$ are consistent with the previous findings of~\citeauthor{Whalley2021-vi}~\cite{Whalley2021-vi}
Overall, we find that the bonding environments for all defects are same for bulk and surface except for iodine interstitials in the +1 charge state. 

\begin{figure}[htbp]
    \centering
    \includegraphics[width=\textwidth]{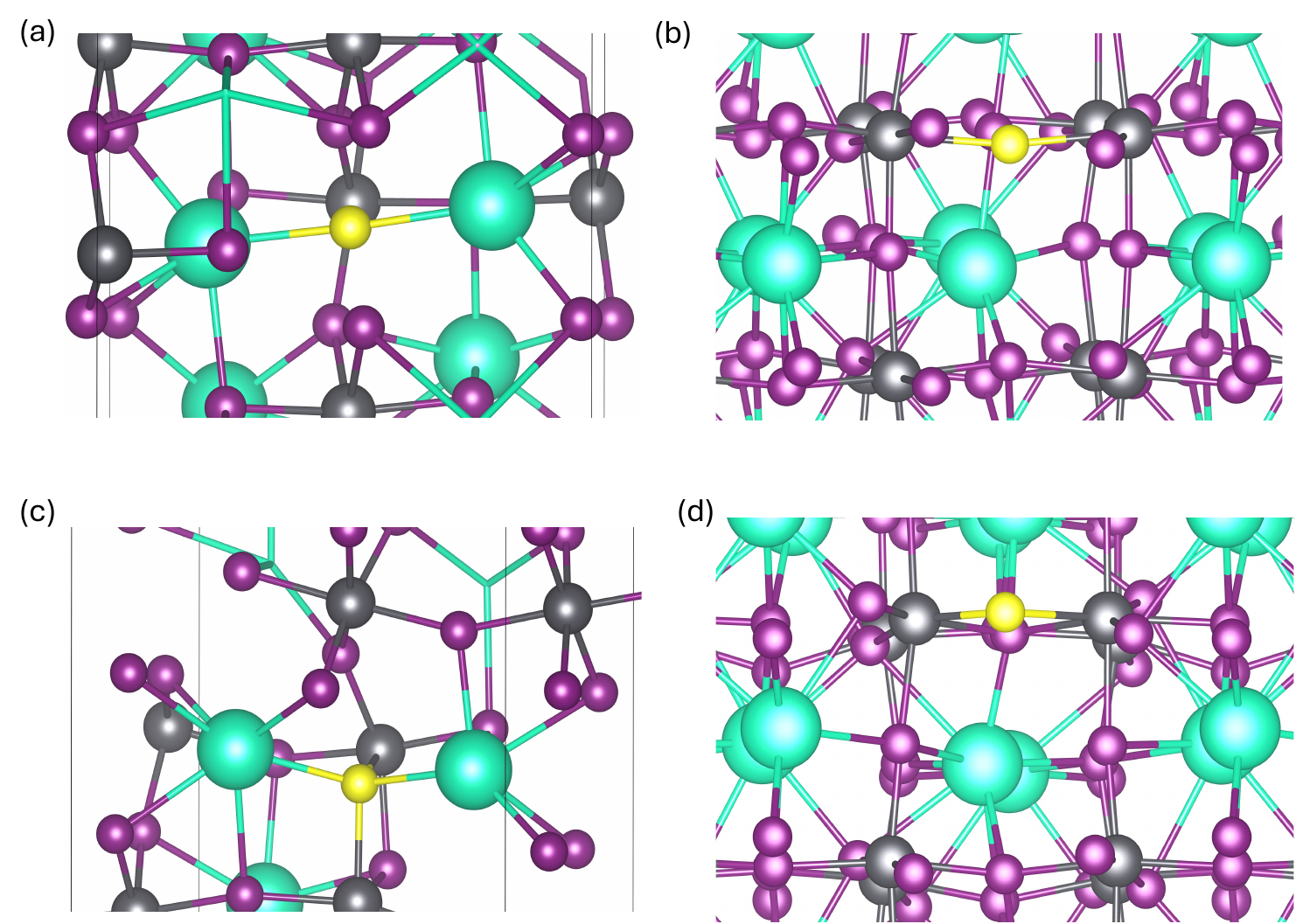}
    \caption{Comparison of interstitial defect configurations in CsPbI$_{3}$ in the bulk and at the surface. The defects are located the in second layer of the 17-layer slab. The yellow atom represents the iodine interstitial. The configurations depicted include (a) $\mathrm{I}_\mathrm{i}^\bullet$ in the bulk, (b) $\mathrm{I}_\mathrm{i}^\bullet$ at the surface, (c) $\mathrm{I}_{\mathrm{i}}^{{\displaystyle \prime }}$ in the bulk, and (d) $\mathrm{I}_{\mathrm{i}}^{{\displaystyle \prime }}$ at the surface. In the bulk, $\mathrm{I}_\mathrm{i}^\bullet$ is positioned between two cesium atoms whereas at the surface, it is found between two lead atoms. For $\mathrm{I}_{\mathrm{i}}^{{\displaystyle \prime }}$ in the bulk, the defect exists as a split interstitial bonded to both cesium and lead atoms, while at the surface, it forms a split interstitial without any bonds with the cesium atoms, highlighting the impact of spatial placement on defect stabilization and bonding within CsPbI$_{3}$.}
    \label{fig:CsPbI3_structures.pdf}
\end{figure}

In contrast to MAPI, we found significant differences in the stable defect configurations of iodine interstitials in the bulk and near surface for \ce{CsPbI3}. The iodine interstitial forms bonds with Cs atoms in the bulk (I-Cs distance = 3.6 and 3.7 for -1 and +1 charge states) but not near the surface (I-Cs distance = 6.0 and 4.9 {\AA} for -1 and +1 charge states). 
Further, as illustrated in \autoref{fig:CsPbI3_structures.pdf}, the $\mathrm{I}_\mathrm{i}^{\bullet}$ in the bulk (panel a) prefers to form bonds with Cs atoms while the $\mathrm{I}_{\mathrm{i}}^{{\displaystyle \prime }}$ in the bulk (panel c) prefers to bond with cesium and lead atoms forming split interstitial in \ce{PbI2} layer consistent with the previous findings of \citealt{doi:10.1021/acs.jpcc.1c08741}. Near the surface, the interstitial iodine only forms bonds with lead atoms (panels b and d). 
In addition, the defect in the -1 charge state exists as a split interstitial at the surface (panel d).
On the other hand, no significant differences were observed in defect configuration between the bulk and surface for iodine vacancies (Fig. S1).  

These findings highlight the differences in the bonding environments between bulk and surface defects. We note that the differences in the properties of defects in the bulk and surfaces can be attributed to three main factors: differences in 1) bonding or coordination environments, 2) bond lengths, and 3) loss or periodicity or presence of vacuum near the surface.

\subsection{Formation energy of bulk versus surface defects}
\begin{figure}[htbp]
    \centering
    \includegraphics[width=\textwidth]{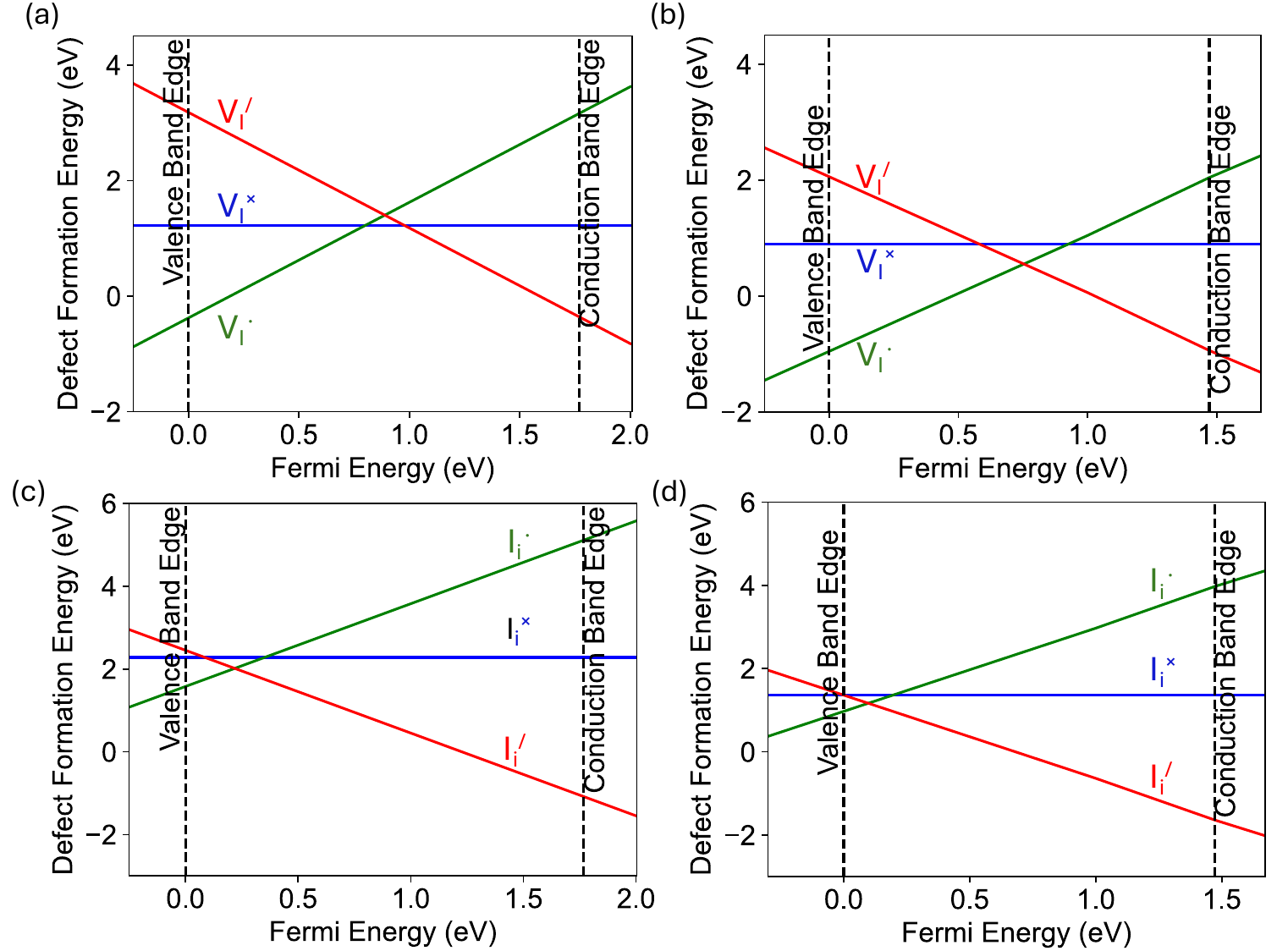}
    \caption{Comparison of DFE plotted against Fermi energy in the bulk and at the surface for MAPI. %
   The DFE is plotted for (a) iodine vacancies in the bulk, (b) surface iodine vacancies, (c) iodine interstitials in the bulk, and (d) surface iodine interstitials. The DFE for both iodine vacancies and interstitials reduces at the surface as compared to bulk.
    }
    \label{fig:MAPI_DFE_vs_Fermi.pdf}
\end{figure}

\autoref{fig:MAPI_DFE_vs_Fermi.pdf} compares the DFE of bulk and surface defects in MAPI as a function of the Fermi energy under the Pb rich and \ce{PbI2} excess conditions employed in this study.
The surface defects considered in the plot are those in the first layer of the slab.  For the vacancy defect, we find that the DFE reduces from the bulk (a) to the surface (b) across all three charge states. For example, the DFE of the vacancy with charge +1 reduces from -0.37 to -0.95 eV at Fermi energy equal to the VBM. %
In addition, the neutral defect which is stable near the middle of the gap for the bulk, is not stable near the surface for any Fermi energy value. The (+/0) and (0/-) transition levels change from 0.80 and 0.98 eV in the bulk to 0.93 and 0.58 eV at the surface, respectively. %
For interstitial defects, the reduction in DFE is slightly higher; 
the DFE for charge state +1  reduces from 1.57 eV in the bulk to 0.97 eV at the surface at the VBM.

\begin{figure}[htbp]
    \centering
    \includegraphics[width=\textwidth]{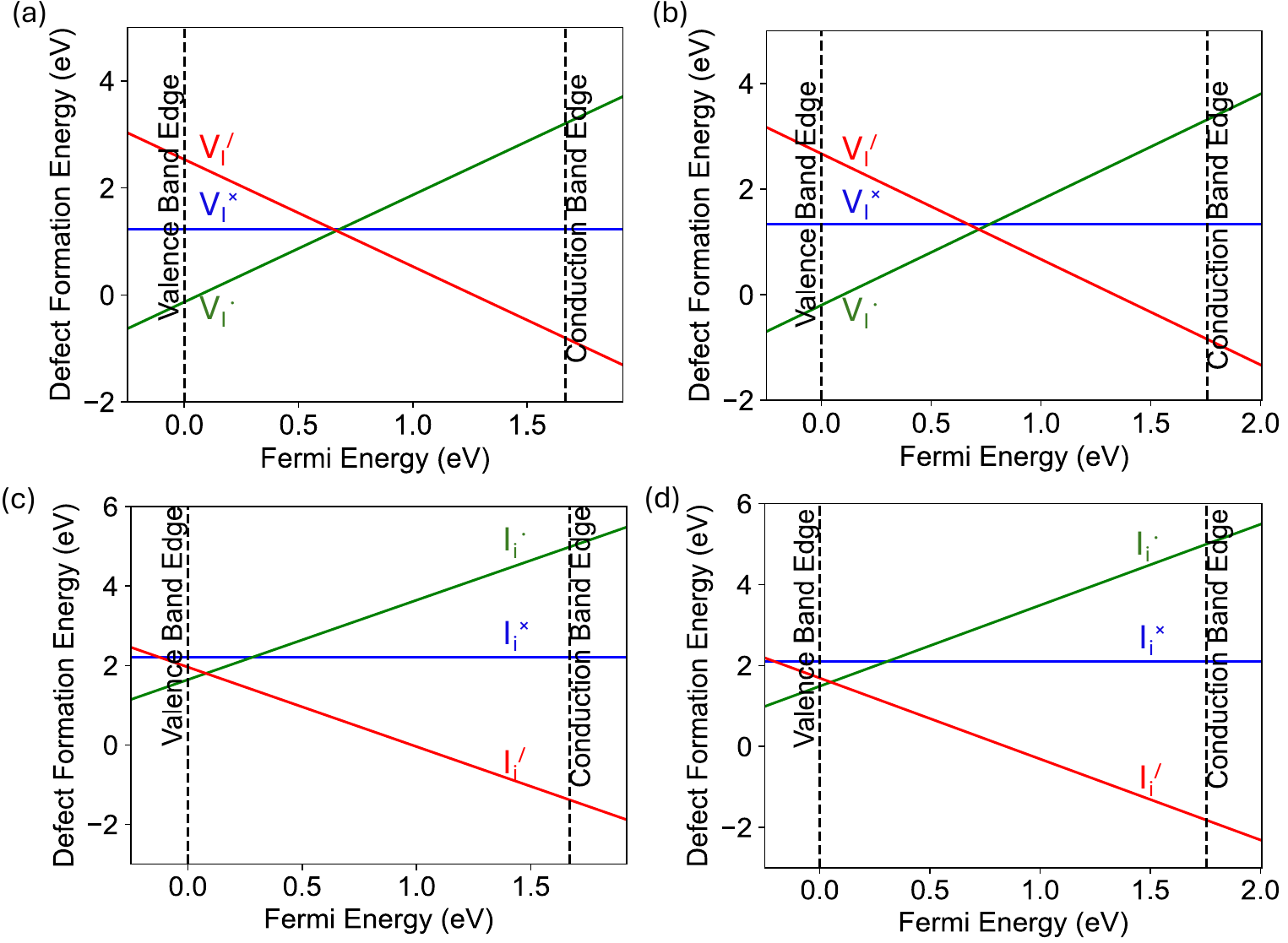}
    \caption{Comparison of variation of DFE with Fermi energy in the bulk and on the surface for CsPbI$_{3}$.
    This figure shows the DFE variation for (a) iodine vacancies in the bulk, (b) iodine vacancies at the surface , (c) iodine interstitials in the bulk, and (d) iodine interstitials at the surface. DFEs at the surface for both vacancy and interstitial defects align closely with bulk values, suggesting similar stability across bulk and surface.}
    \label{fig:CsPbI3_DFE_vs_Fermi.pdf}
\end{figure}

The DFEs in the bulk and at the surface of CsPbI$_{3}$ are compared in \autoref{fig:CsPbI3_DFE_vs_Fermi.pdf}. In contrast to MAPI, we find that differences between bulk and surface defects are smaller for both vacancies and interstitials. For example, the DFE changes by only 0.07 eV for the positively charged iodine vacancy at the surface compared to 0.58 eV for MAPI at the VBM.

The above analysis examines  the differences in the DFEs in the  bulk and at the surface of MHPs. This characterization sets the stage for a more detailed analysis in the next section, where we consider how these DFEs vary as a function of distance from the surface. By examining this variation across multiple layers of MHP structures, we provide insights into the depth-dependent behavior of defects and the transition from surface-like to bulk-like behavior. 
This spatial analysis is critical for understanding how defects are distributed over the thickness of MHPs, which has direct implications for designing effective passivation strategies for enhanced device performance.

\subsection{Variation of defect formation energy with distance from the surface}

\begin{figure}[htbp]
    \centering
    \includegraphics[width=\textwidth]{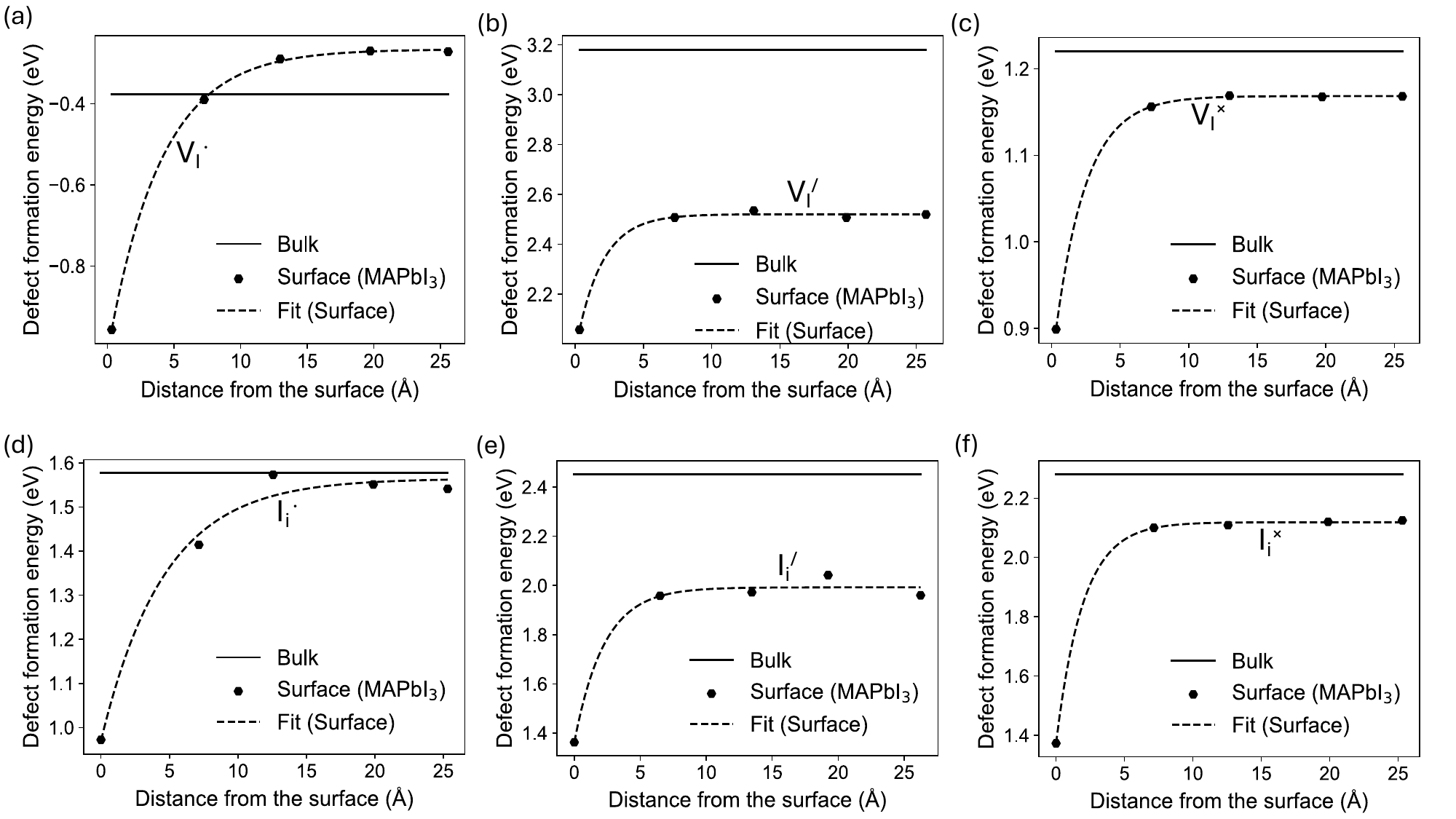}
    \caption{DFE as a function of distance from the surface in a 17-layer MAPI structure. This figure illustrates the variation of DFE for (a) $\mathrm{V}_{\mathrm{I}}^\bullet$, (b) $\mathrm{V}_{\mathrm{I}}^{{\displaystyle \prime }}$, (c) $\mathrm{V}_{\mathrm{I}}^{\times}$, (d) $\mathrm{I}_\mathrm{i}^\bullet$, (e) $\mathrm{I}_{\mathrm{i}}^{{\displaystyle \prime }}$, and (f) $\mathrm{I}_{\mathrm{i}}^{\times}$. Vacancy defects in all charge states show significant reductions in surface DFE compared to the bulk, suggesting a higher likelihood of vacancy defect formation at the surface. Similarly, for interstitial defects, the DFE values are lower at the surface, indicating their increased probability to form near the surface. The exponential trend of increasing DFE with distance from the surface is observed in all plots, highlighting the complex impact of surface proximity on defect stability. }
    \label{fig:MAPI_DFE_vs_DIstance.pdf}
\end{figure}

To understand the variation of DFE with the distance from the surface, we employ a 17-layer MAPI slab, which consists of alternating layers of MAI and PbI$_{2}$ as shown in Fig. S2. Defects situated in the deeper layers of the slab will exhibit properties similar to those of the bulk material.
The DFE for iodine vacancies and interstitials in three different charge states at $\varepsilon_F=0$ is plotted in \autoref{fig:MAPI_DFE_vs_DIstance.pdf}. In all cases, we find that the DFE is lower at the surface (first layer) compared to the bulk, with the difference ranging from 0.32 eV for $\mathrm{V}_{\mathrm{I}}^{\times}$ to as large as 1.12 eV for $\mathrm{V}_{\mathrm{I}}^{{\displaystyle \prime }}$. The difference between bulk and surface DFEs is typically smaller for neutral defects than charged defects. Further, our layer-dependent calculations elucidate how the DFE transitions from surface to the bulk value. The DFE follows a saturating exponential behavior, eventually converging to a fixed value. However, we observe discrepancies between the converged DFE and the bulk values, spanning from 0.05 to 0.66 eV.
These differences may arise from variations in relaxed bond lengths or bonding environments between bulk systems and defects situated in deeper layers within surface slabs used for DFT calculations.  These discrepancies will be discussed in the next section. Additionally, errors introduced due to charge correction schemes may also contribute to the discrepancy.  

\begin{figure}[htbp]
    \centering
    \includegraphics[width=\textwidth]{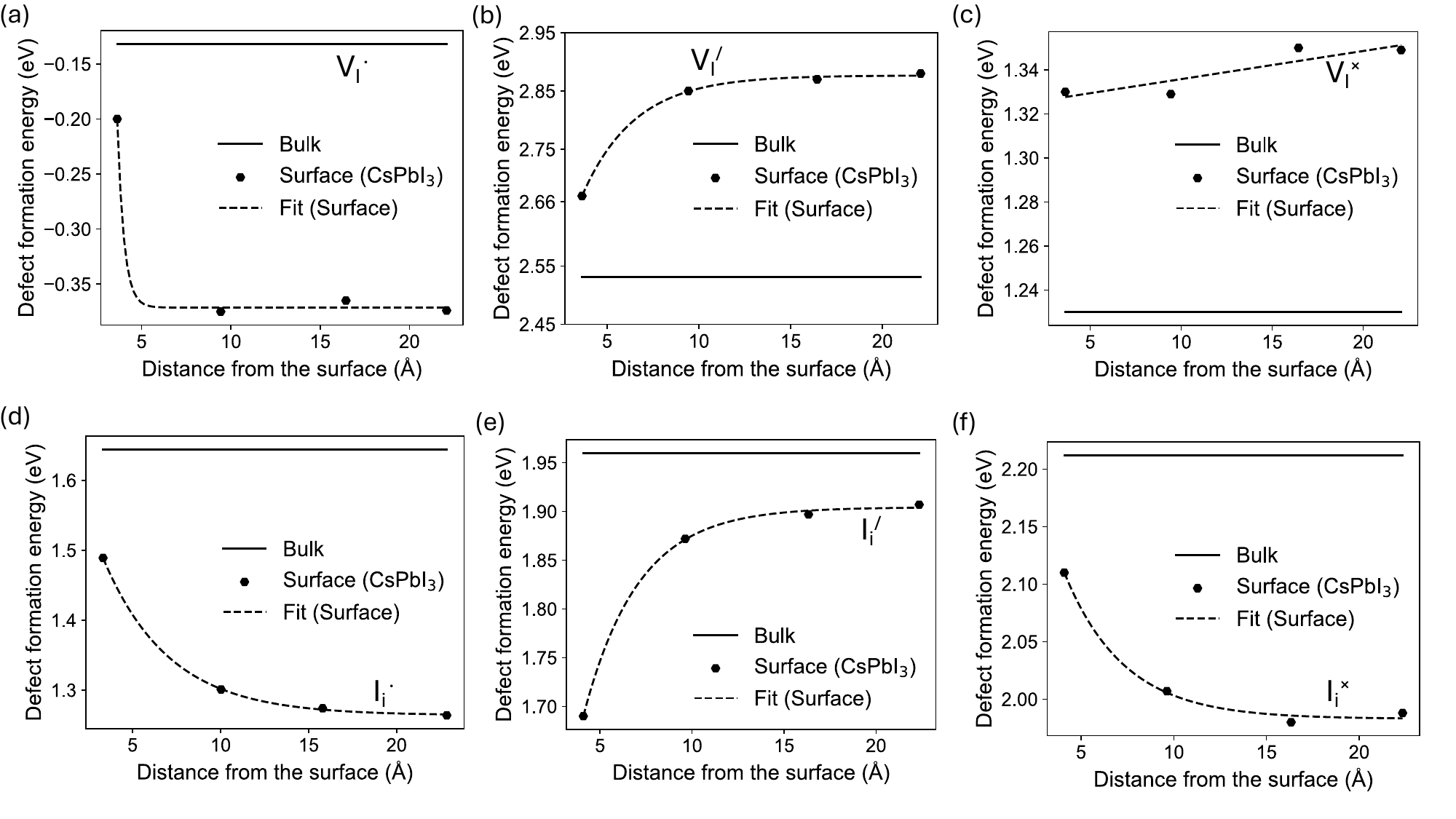}
    \caption{Defect Formation Energy (DFE) as a function of distance from the surface in a 17-layer \ce{CsPbI3} slab  for (a) $\mathrm{V}_{\mathrm{I}}^\bullet$, (b) $\mathrm{V}_{\mathrm{I}}^{{\displaystyle \prime }}$, (c) $\mathrm{V}_{\mathrm{I}}^{\times}$, (d)  $\mathrm{I}_\mathrm{i}^\bullet$, (e) $\mathrm{I}_{\mathrm{i}}^{{\displaystyle \prime }}$, and (f) $\mathrm{I}_{\mathrm{i}}^{\times}$. (a) shows reduction in surface DFE, indicating a preference for defect formation at the surface. (b) and (c) show higher surface DFE as compraed to bulk reducing the chances of defect formation at the surface. (d) through (f) illustrate that surface DFEs for interstitial defects are consistently lower than those in the bulk, suggesting a higher likelihood of defect formation at the surface. }
    \label{fig:CsPbI3_DFE_vs_distance.pdf}
\end{figure} 
The variation in DFE for iodine vacancies and interstitials in CsPbI$_3$ slab is plotted in \autoref{fig:CsPbI3_DFE_vs_distance.pdf}. While the observed saturating exponential behavior of DFEs is similar to that of MAPI, we identify three key differences. First, the differences between bulk and surface DFEs are relatively modest~\cite{D3CP00285C}, ranging from 0.07 eV for $\mathrm{V}_{\mathrm{I}}^\bullet$ to 0.27 eV for $\mathrm{I}_{\mathrm{i}}^{{\displaystyle \prime }}$.  
Second, we observe a reversal in the DFE trend where the slab DFE moves away from the bulk DFE for all the cases except for $\mathrm{I}_{\mathrm{i}}^{{\displaystyle \prime }}$.
Finally, the surface DFE is larger than the bulk DFE for $\mathrm{V}_{\mathrm{I}}^{\times}$ and $\mathrm{V}_{\mathrm{I}}^{{\displaystyle \prime }}$. 
The deviation may be attributed to the different stable configurations of the defects and different bond lengths surrounding the defects, resulting in distinct bonding environments.

The differences between bulk and surface DFEs will result in highly inhomogeneous defect densities in MHPs. For example, a $\Delta E = 0.32$ eV difference between bulk and surface DFEs observed for $\mathrm{V}_{\mathrm{I}}^{\times}$ in MAPI can result in an exponential increase in defect density at the surface, approximately $\exp\left(  \Delta E / k_B T\right)\approx 10^{5}$ times higher than in the bulk together with increased defect-defect interactions. Greater inhomogeneity may arise for higher $\Delta E$ values predicted for other defects.  Our findings are similar to previous results predicting a millionfold increase in defect density at the interface between cubic and hexagonal polytypes of \ce{CsPbI3}~\cite{wooInhomogeneousDefectDistribution2023}.

Our calculations extend beyond surface defects and provide a more comprehensive understanding by mapping DFEs as a function of distance from the surface, revealing complete spatial distribution and length of the transition region. For both MHPs, we fitted the variation in DFE with distance  to  the saturating exponential function, 
\begin{equation}\label{eq:dfeexp}
\text{DFE}(x) = a + b \exp\left(-\frac{x}{l}\right)
\end{equation}
where \textit{x} is the distance of defect from the surface and \textit{a}, \textit{b}, and \textit{l} are the constants. This function helps in quantifying the defect behavior at the surface and transition regions. The parameter $l$ represents the length scale for the decay in surface effects and is hereafter referred to as the decay length.

The decay lengths for positively and negatively charged iodine vacancies and interstitials in MAPI and \ce{CsPbI3} are tabulated in \autoref{tab:my_label}.
The values of decay length for MAPI were relatively small, ranging from 0.32 {\AA} to 0.77 {\AA}. This indicates a rapid decrease in surface effects, suggesting that the DFE values converge to bulk values within a small distance from the surface. In MAPI, the impact of the surface on defect energetics is limited to 1 or 2 layers, which might be beneficial for situations where surface defects are unwanted. CsPbI$_3$ exhibits a range of decay lengths, the highest being 3.75 {\AA} for $\mathrm{I}_\mathrm{i}^\bullet$. The larger decay length implies a gradual convergence towards the DFE in bulk, indicating a prolonged effect of the surface compared to MAPI.  The decay length is an important property to quantify how surfaces, grain boundaries, and interfaces influence the performance of MHP-based devices. For example, a higher decay length suggests a greater need for surface passivation in the deeper layers to control the high defect density in subsurface layers of the material. The larger range of interaction influences the rates at which carriers recombine,  the migration of ions, and thus the stability and efficiency of MHP solar cells.

\begin{table}[htbp]
    \centering
    \begin{tabular}{|c|c|c|c|}
    \hline
\textbf{Defect} & \textbf{\textit{l}({\AA})} & \textbf{Defect} & \textbf{\textit{l}({\AA})} \\ \hline
MAPI $\mathrm{V}_{\mathrm{I}}^\bullet$ & 0.666 & CsPbI$_3$ $\mathrm{V}_{\mathrm{I}}^\bullet$ & 0.062 \\ \hline
MAPI $\mathrm{V}_{\mathrm{I}}^{{\displaystyle \prime }}$ & 0.315 & CsPbI$_3$ $\mathrm{V}_{\mathrm{I}}^{{\displaystyle \prime }}$ & 0.477 \\ \hline
MAPI $\mathrm{I}_\mathrm{i}^{\bullet}$ & 0.770 & CsPbI$_3$ $\mathrm{I}_\mathrm{i}^{\bullet}$ & 3.748 \\ \hline
MAPI $\mathrm{I}_{\mathrm{i}}^{{\displaystyle \prime }}$ & 0.376 & CsPbI$_3$ $\mathrm{I}_{\mathrm{i}}^{{\displaystyle \prime }}$ & 0.493 \\ \hline
    \end{tabular}
    \caption{Values of decay length for the function DFE(x) [\autoref{eq:dfeexp}] for iodine vacancies and interstitials in MAPI and CsPbI$_3$.}
    \label{tab:my_label}
\end{table}

\subsection{Analysis of bond lengths in defective structures}
\begin{figure}[htbp]
    \centering
    \includegraphics[width=\textwidth]{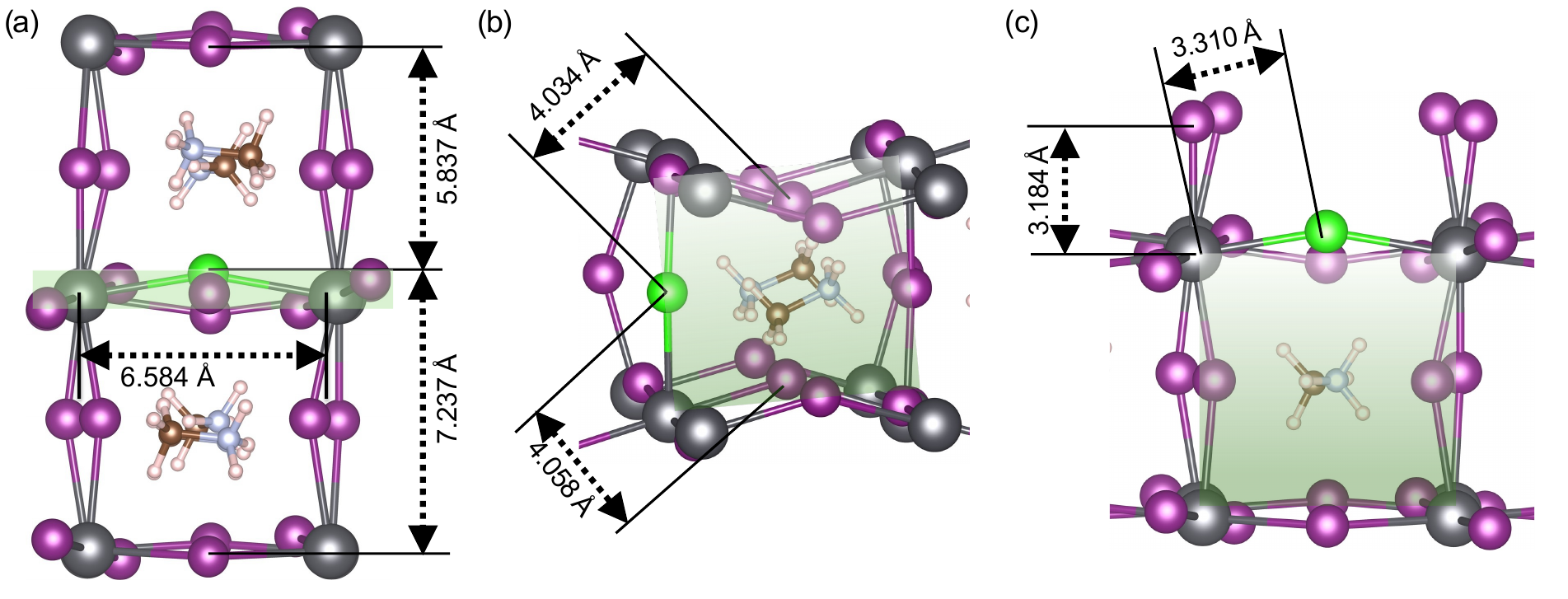}
    \caption{ Visualization of Pb-Pb distances, Pb-I bond lengths, and I-I distances on MAPI surface for $\mathrm{I}_{\mathrm{i}}^\bullet$ examined to explain the variation in DFE with distance.  (a) out of plane I-I distances, and Pb-Pb distances, (b) in-plane I-I distances, and (c) Pb-I bond lengths. The green-colored atom represents interstitial iodine. The shaded green plane represents the Pb-I layer where the defect is located.}
    \label{fig:S_1.pdf}
\end{figure}

To explain the variation in DFE with the location of defects in different layers, we conducted a comprehensive analysis of bond lengths and distances between atoms near the defect. Specifically, we examined the variation in Pb-Pb distances, Pb-I bond lengths, and I-I distances with the distance of the defect from the surface. 
\autoref{fig:S_1.pdf}  shows the bonds and atomic distances considered for this analysis in the case of MAPI: Pb-Pb distances, out of plane I-I distances, in-plane I-I distances, and Pb-I bond lengths surrounding the defect. Note that the Pb-Pb distances were considered for those lead atoms that were bridged by the iodine atom removed during vacancy formation. Also, the in-plane distances lie in the plane of the layer, whereas the out-of-plane distances represent the distances perpendicular to the layer. For vacancy defects, we analyze the Pb-Pb distances to explain the variation in DFE while for interstitial defects, we examine the Pb-I bond lengths surrounding the defect.

\begin{figure}[htbp]
    \centering
    \includegraphics[width=\textwidth]{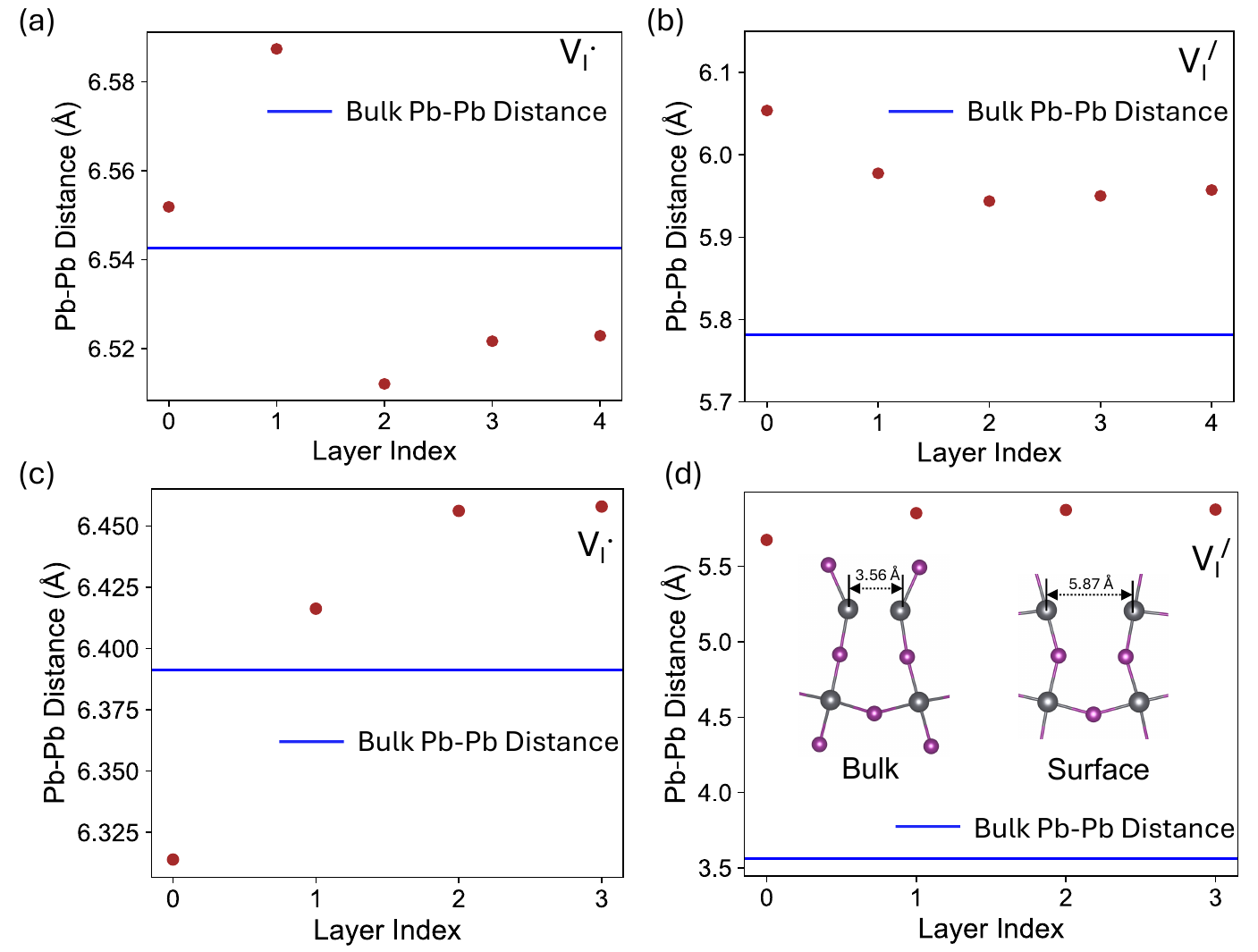}
    \caption{Pb-Pb distances in MAPI and \ce{CsPbI3} containing vacancy defects plotted against the layer index (0=surface layer) %
    for (a) $\mathrm{V}_{\mathrm{I}}^\bullet$ in MAPI, (b) $\mathrm{V}_{\mathrm{I}}^{{\displaystyle \prime }}$ in MAPI, (c) $\mathrm{V}_\mathrm{I}^\bullet$ in \ce{CsPbI3}, and (d) $\mathrm{V}_{\mathrm{I}}^{{\displaystyle \prime }}$ in \ce{CsPbI3}, respectively. These variations in Pb-Pb distances result in convergence of DFE towards or away from bulk values. The inset in (d) shows the surprisingly larger difference in Pb-Pb distance between the bulk and deepest layer of the slab. The brown circles represent the Pb-Pb distances in the slabs while the bulk distance is represented by the blue horizontal line.}
    \label{fig:MAPI_Pb-Pb_BL.pdf}
\end{figure}

When an iodine vacancy is formed, the lead atoms bridged by the removed iodine atom experience a repulsion that is dependent on their charge state~\cite{wooInhomogeneousDefectDistribution2023}. Hence, lead atoms rearrange in response and a larger Pb-Pb distance corresponds to a lower energy. \autoref{fig:MAPI_Pb-Pb_BL.pdf} compares the bulk and slab Pb-Pb distances when the vacancy is located in different layers versus the layer index for MAPI and \ce{CsPbI3}.
We note that in all cases, the Pb-Pb distances converge leading to the convergence of the slab DFE with an increasing number of layers. In addition, the difference between the converged slab DFE and the bulk DFE can also be explained by the difference in the Pb-Pb distances except (d). For example, in (a) the converged Pb-Pb distance is lower in the slab compared to the bulk, leading to more repulsion and higher converged DFE for the slab. For $\mathrm{V}_{\mathrm{I}}^{{\displaystyle \prime }}$ in \ce{CsPbI3} (d), the converged Pb-Pb distances in the slab are significantly higher than the bulk, yet the bulk DFE is lower than the slab. As seen in the inset of (d), the Pb-Pb distance in bulk is only 3.56 {\AA} while in the deepest layer, it is 5.87 {\AA}. We believe that other factors such as Pb-I bond lengths (Fig S5) and different charge distribution on lead atoms might be determining the DFE for this case. Further, we also plot the variation of in-plane and out of plane I-I distances in the bulk and slab in Fig. S3 and S4 for MAPI and \ce{CsPbI3} respectively.

\begin{figure}[htbp]
    \centering
    \includegraphics[width=\textwidth]{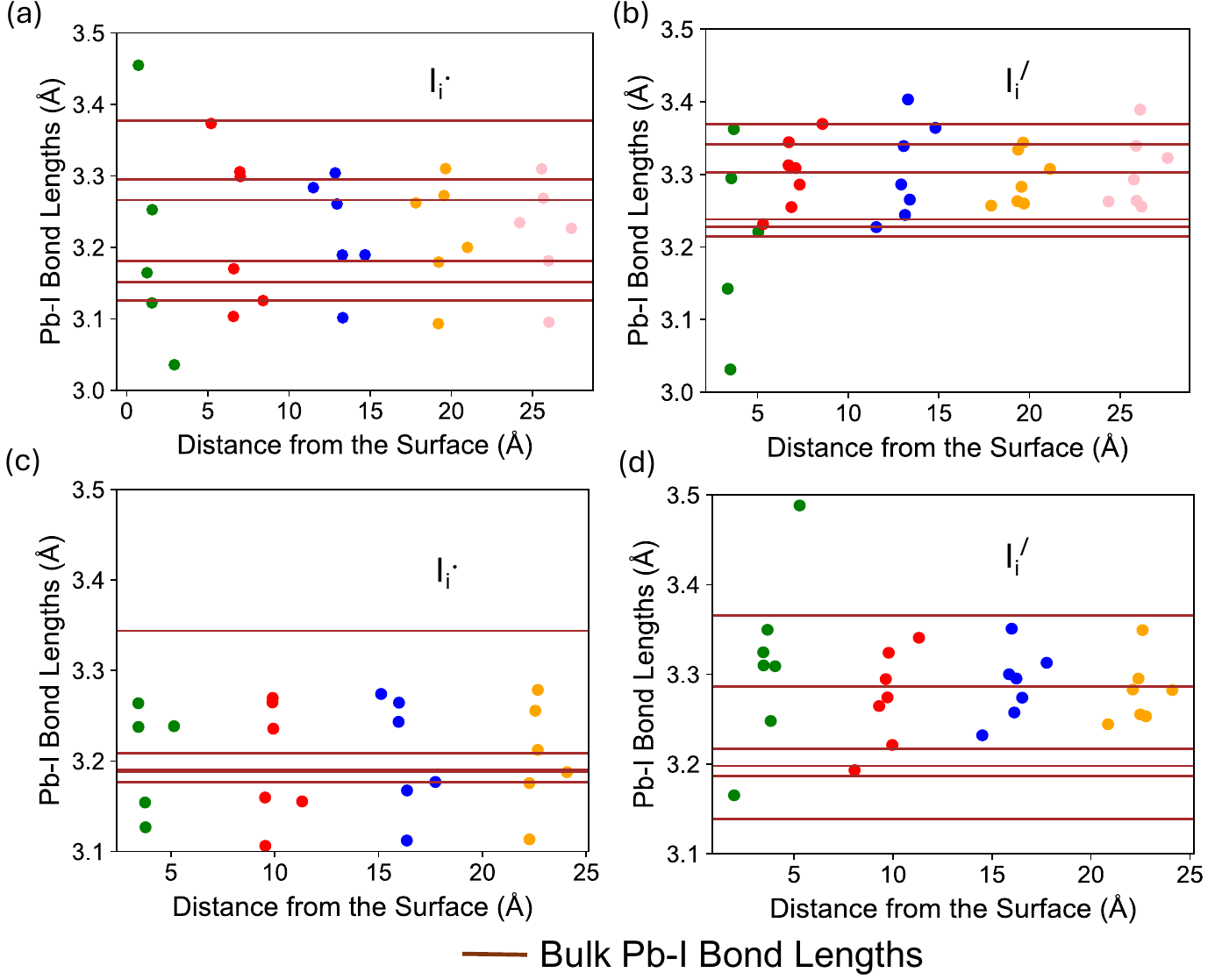}
    \caption{Pb-I bond lengths in MAPI and \ce{CsPbI3} present in the neighborhood of interstitial defects plotted for different layers against the average distance of lead and iodine atoms from the surface. The plots show the variation in Pb-I bond lengths when the defect is located in different layers 
    for (a) $\mathrm{I}_{\mathrm{i}}^\bullet$ in MAPI, (b) $\mathrm{I}_{\mathrm{i}}^{{\displaystyle \prime }}$ in MAPI, (c) $\mathrm{I}_\mathrm{i}^\bullet$ in \ce{CsPbI3}, and (d) $\mathrm{I}_{\mathrm{i}}^{{\displaystyle \prime }}$ in \ce{CsPbI3}, respectively.  The circles represent the slab bond lengths while the horizontal lines denote bulk bond lengths. Pb-I bond lengths in different layers are denoted by different colors. These converging Pb-I bond lengths towards the bulk values result in the convergence of DFE towards bulk values with distance from the surface.}
    \label{fig:MAPI_Pb-I_BL.pdf}
\end{figure}

For interstitial defects, we plotted the Pb-I bond lengths in the defective structure when the defect is located in different layers of the slab in \autoref{fig:MAPI_Pb-I_BL.pdf} for both MAPI and \ce{CsPbI3}. Only Pb-I bond lengths in the neighborhood of the defect are plotted. For each bond, the bond length is plotted against the average distance of the lead and iodine atoms from the surface. The bulk Pb-I bond lengths are also shown as horizontal lines for comparison. 
We find that the deviation in slab Pb-I bond lengths from the bulk is significantly higher in \ce{CsPbI3} compared to MAPI. This deviation occurs due to different stable configurations of the iodine interstitial in the slab and the bulk for \ce{CsPbI3} as discussed earlier. We also find that except for (c), all Pb-I bond lengths move closer towards bulk bond lengths, leading to convergence of DFE towards the bulk value in panels a, b, and d. In (c), the  Pb-I bond lengths consistently differ from the bulk values across all layers, leading to a large difference between the converged slab DFE and the bulk value.
We also plot the variation of I-I distances, Pb-MA, and Pb-Cs distances with defect location in Fig. S3, S4, S6 and S7.

\subsection{Variation of defect transition levels with distance from the surface}

\begin{figure}
    \centering
    \includegraphics[width=\textwidth]{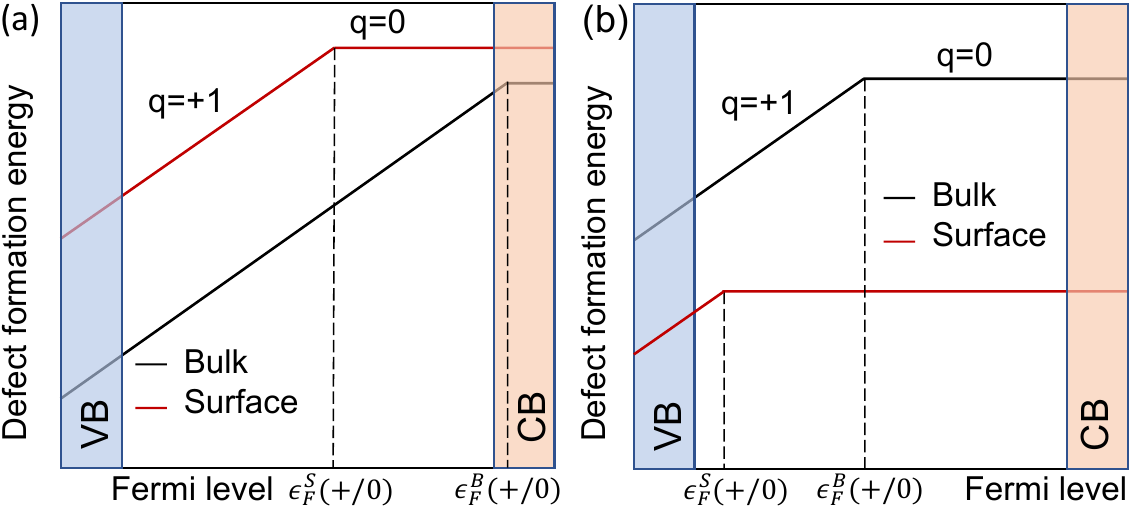}
    \caption{Differences in defect transition level between the bulk, $\epsilon_F^B(+/0)$ and at the surface, $\epsilon_F^S(+/0)$. (a) A shallow defect in the bulk becomes a deep defect at the surface and (b) a deep defect in the bulk is converted to a shallow defect by the surface.}
    \label{fig:transition}
\end{figure}

The variation in DFE as a function of distance from the surface will also lead to corresponding changes in defect transition levels of different layers, thereby affecting non-radiative recombination rates.  An extreme scenario is sketched in \autoref{fig:transition} where the surface completely changes the trapping nature of the defect. In panel (a), a shallow bulk defect is transformed into a deep surface defect due to the modified DFE at the surface. The (+1/0) transition level changes from a value within the conduction band to nearly the middle of the gap. In panel (b), a deep bulk defect is rendered shallow at the surface. Such transformations of the character of defect traps at the surface may be detrimental for photovoltaic and optoelectronic applications, hence, worth investigating for MHPs. Furthermore, these transformations may also result in a depletion of charge carriers at the surface and an accumulation in the bulk, when a shallow bulk trap becomes a deep surface trap, leading to a redistribution of  charge carriers.

\begin{figure}[htbp]
    \centering
    \includegraphics[width=\textwidth]{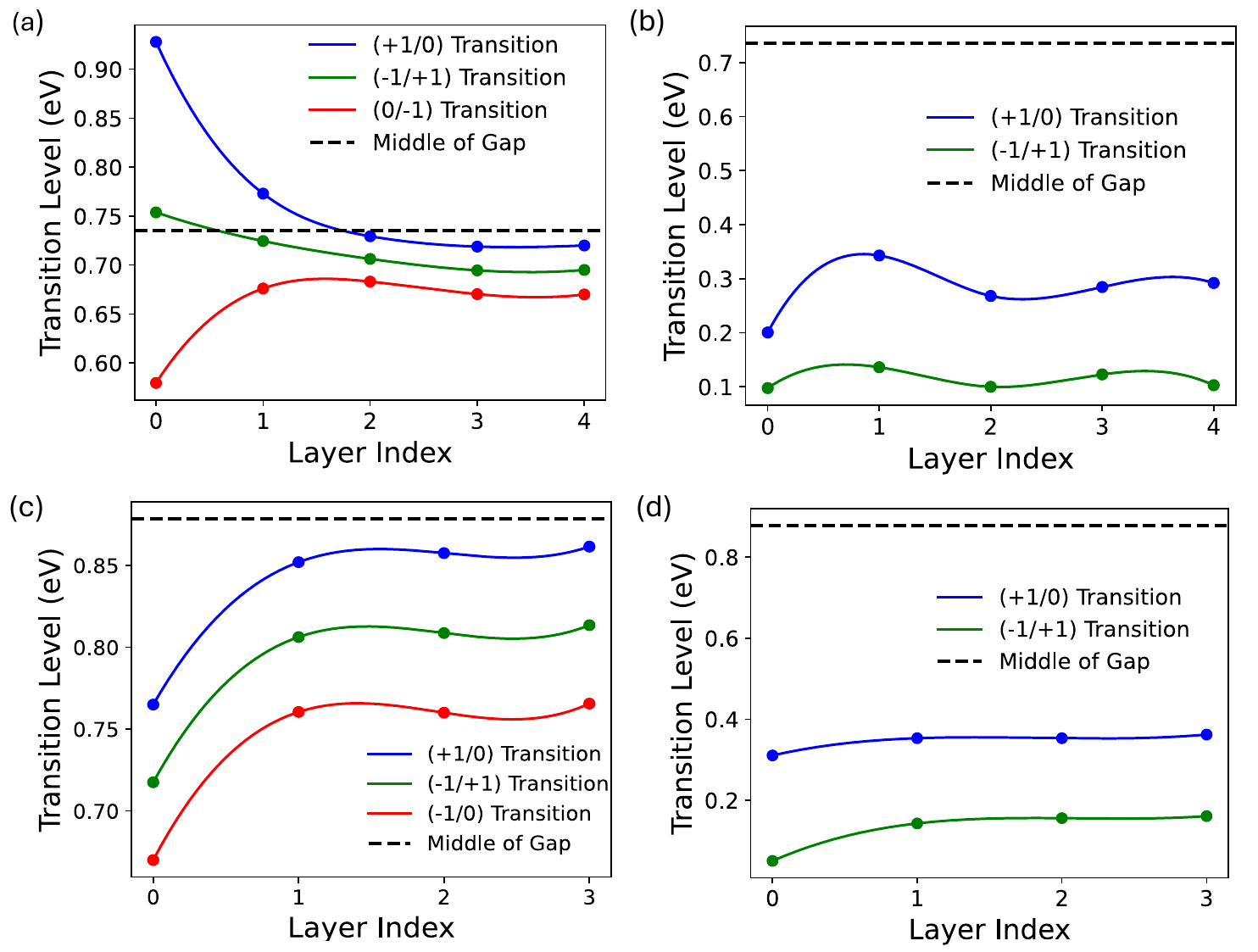}
    \caption{Defect transition levels in  MAPI and CsPbI$_3$ as a function of layer index away from the surface (0=surface layer) for  (a) vacancy defect in MAPI, (b) interstitials defects in MAPI, (c) vacancy defects in \ce{CsPbI3}, and (d) interstitial defects in \ce{CsPbI3}. Both materials exhibit changes in defect transition levels with layer depth, emphasizing the implications for charge carrier dynamics and on the efficiency of MHP-based photovoltaic devices.}
    \label{fig:structure}
\end{figure}

We calculate the defect transition levels using the DFEs of iodine vacancies and interstitials in three charge states predicted for different layers of MHPs and plot them as a function of layer index in \autoref{fig:structure}. The dashed black line denotes the middle of the gap as a reference for assessing whether the defect is a deep or shallow trap. In both MAPI and \ce{CsPbI3}, the transition levels for iodine vacancies shift towards the middle of the gap as the defect moves away from the surface and deeper into the bulk. The change in transition levels of vacancies from the surface to the deepest layer considered in our study is $\sim 0.05-0.2$ eV for MAPI and $\sim 0.09-0.1$ eV for \ce{CsPbI3}.
For iodine interstitials, we calculated only (+1/0) and (-1/+1) transition levels since the transition level for (-1/0) was below the valence band energy.
The changes in transition levels of iodine interstitials are more modest, in the range $\sim 0.00-0.09$ eV for MAPI and $\sim 0.05-0.11$ eV for \ce{CsPbI3}.

Our layer-dependent analysis points to iodine vacancies becoming deep traps on moving away from the surface, leading to increased non-radiative recombination. While the exact transition level and non-radiative recombination rate will depend on various conditions such as chemical potentials, phonon emission routes, etc., this analysis outlines the need for understanding the transition in the nature of defects as traps for thin film MHP solar cells. The knowledge obtained from this analysis is crucial in guiding the synthesis and treatment of MHPs to reduce harmful deep-level traps that may exist in the surface and subsurface regions to improve the PCE of MHP solar cells.

\section{Summary and Conclusions}
By performing first-principles calculations on defects in MHPs at varying depths from the surface, we have uncovered unique insights into how their properties are modulated near surfaces. The differences in DFEs in the bulk and surfaces will lead to defect enrichment or depletion in certain regions, affecting the surface recombination velocity and charge carrier dynamics. Further, the saturating exponential behavior of DFE obtained from our calculations can be used to extract complete defect density profiles in MHPs with varying thicknesses, grain sizes, and interfaces. 
The decay lengths extracted from the DFE versus distance plots quantify the spatial extent of the influence of surface on defect properties. We attribute the changes in DFE from the surface to the bulk to differences in bond lengths, notably Pb-Pb and Pb-I bond lengths near the defects and different bonding environments. Finally, we find that the charge transition levels for defects can also change near surfaces by $\sim$ 0.4 eV compared to the bulk for MHPs, leading to changes in their trapping tendency.

Our investigation paves the way for future studies to explore how we can tailor DFE, defect density, transition level profiles, and decay lengths to minimize non-radiative losses in MHPs.  Designer interfaces, doping, surface, and subsurface passivation techniques may be employed to modify the profiles. Depth-dependent manipulation strategies have shown significant improvements in device efficiency by concurrently modulating bulk and interfacial defects, enhancing charge transport and reducing non-radiative losses.~\cite{D1EE02287C} Also, chemical polishing combined with sub-surface passivation has been shown to effectively reduce defects and enhance electrical contact with the hole transport layer leading to improved PCE of MHP solar cells.~\cite{D1EE00062D}. 
Additionally, our study points out the need for accurate treatment of correction schemes for charged defects at surfaces and interfaces, as these values are usually higher than for defects in bulk.

\begin{acknowledgement}
BA acknowledges the Distinguished Graduate Student Award from Texas Tech University for support of this research.
We also acknowledge the High-Performance Computing Center (HPCC) at Texas Tech University and the Lonestar6 research allocation (DMR23017) at the Texas Advanced Computing Center (TACC) for providing computational resources that have contributed to the research results reported in this paper.

\end{acknowledgement}
\begin{suppinfo}

Details of computational methods, structural information and additional analyses conducted.

\end{suppinfo}

\bibliography{zotero_refs,refs}

\includepdf[pages=1-10]{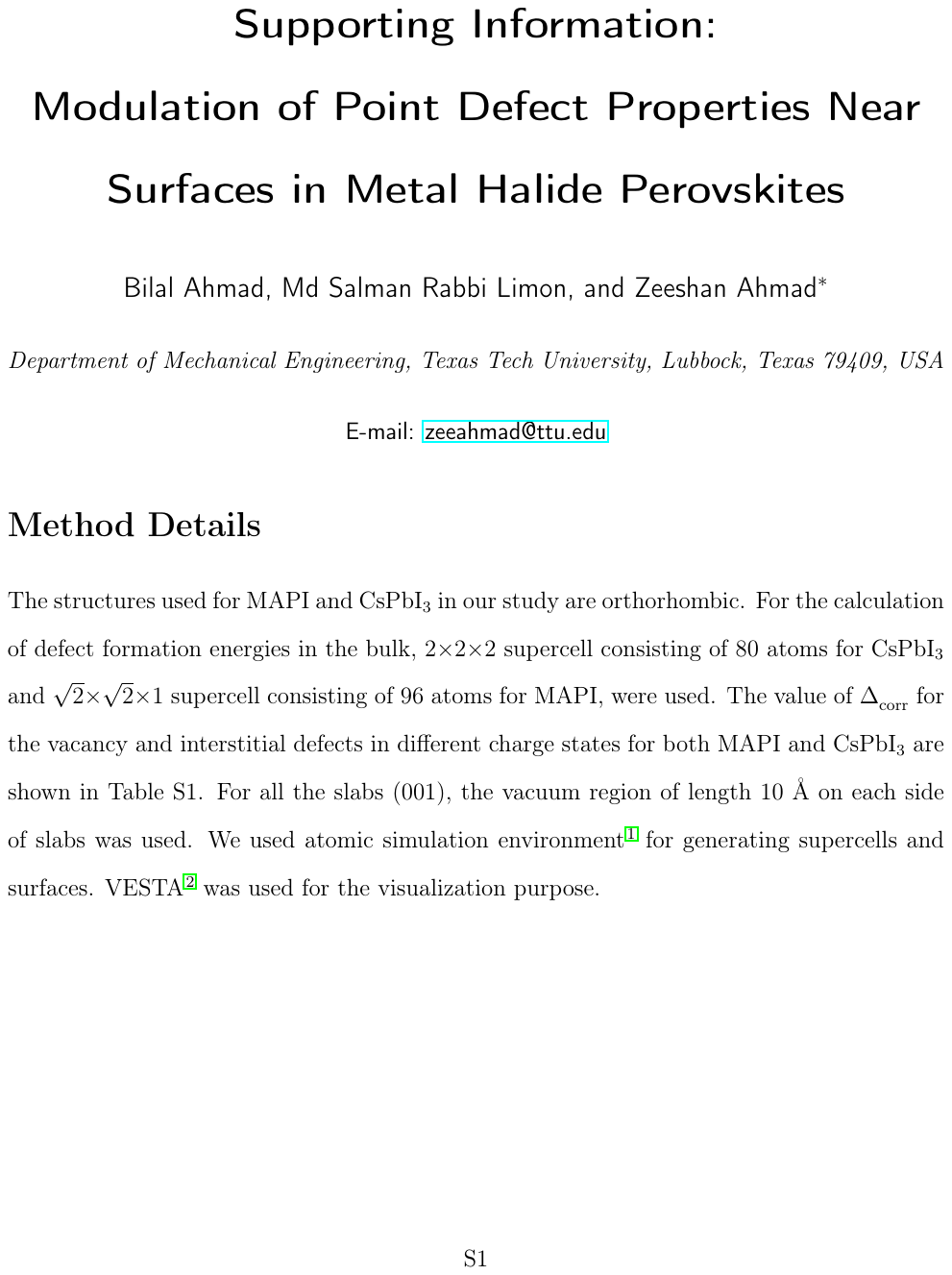}

\end{document}